\documentclass[a4paper, 12pt]{article}

\usepackage{amsmath}
\usepackage{amssymb}
\usepackage[hidelinks]{hyperref}
\usepackage{graphicx}
\begin{document}
	
\title{New method for SISO strong stabilization with advantages over existing methods}
\author{Abdul Hannan Faruqi\thanks{ahannanf20@iitk.ac.in} \and Anindya Chatterjee\thanks{anindya@iitk.ac.in, anindya100@gmail.com}}
\date{Mechanical Engineering, IIT Kanpur, 208016\\ \smallskip \today}

\maketitle
\begin{abstract}
We address stabilization of linear time-invariant (LTI) single-input single-output (SISO) systems in the Laplace domain, with a stable controller in a single feedback loop. Such stabilization is called {\em strong}.
Plants that satisfy a parity interlacing property are known to be strongly stabilizable. Finding such controllers is a well known difficult problem. Existing general methods are based on either manual search or a clever use of Nevanlinna-Pick interpolation with polynomials of possibly high integer order. Here we present a new, simple, and general method for strongly stabilizing systems of relative degree less than 3. We call our method {\em Real to Integers} (RTI). Our theoretical contributions constitute proposing the functional form used, which involves a product of several terms of the form $\displaystyle \left ( \frac{s+a}{s+b} \right )^m$, showing that real $m$'s will arise whenever the plant is strongly stabilizable, and proving that integer $m$'s can be obtained by continuously varying free parameters
(i.e., the $a$'s and $b$'s). Our practical contributions include demonstrating a simple way, based on a trigonometric trick, to adjust the real powers until they take reasonable integer values. We include brief but necessary associated discussion to make the paper accessible to a broad audience. We also present ten numerical examples of successful control design with varying levels of difficulty, including plants whose transfer functions have relative degrees of 0, 1 or 2; and with right half plane zeros of multiplicity possibly exceeding one.\\

\noindent \textbf{Keywords}: LTI, SISO, strong stabilization, control
\end{abstract}

\newtheorem{proof}{Proof}

\newtheorem{point}{Observation}

\newtheorem{cons}{Constraint}

\newtheorem{ex}{Example}

\newtheorem{ct}{Remark}

\newtheorem{pp}{Theorem}

\newtheorem{dfn}{Definition}

\newtheorem{step}{Step}

\newcommand{\bmt}[1]{{\boldmath \mbox{$ #1 $}}}

\newcommand{\drop}[1]{}

\section{Introduction}
\label{sec:intro}
In classical linear time-invariant SISO control with a single feedback loop, a long-standing, difficult problem is that of stabilizing an unstable plant using a stable controller. Such stabilization is called {\em strong}, and it is not possible for every plant. Stabilization with a stable controller is desirable
in many practical situations. Stability of the controller makes starting of the system easier because control cards can be kept offline (with zero input and therefore zero output) until they are warmed up and ready, and then the feedback loop can be engaged with a single switch at a convenient moment when the plant is sufficiently close to the operational condition. In contrast, with an unstable controller, the control card cannot be kept offline because it can develop unbounded control signals even with nominally zero inputs to it; and therefore {\em both} plant and controller have to be brought to a convenient state before the system is released into stable controlled behavior.

As noted above, strong stabilization is not possible for every plant.
Conditions for strong stabilizability of a plant were given by Youla {\em et al.} \cite{youla1974single}. The problem has since been discussed by many authors, including in textbooks by Doyle {\em et al.} \cite{doyle2013feedback} and Vidyasagar \cite{vidyasagar2011control}. Smith and Sondergeld \cite{SMITH1986127} have examined the minimum order required for such a strongly stabilizing controller. G\"unde\c{s} and \"Ozbay \cite{GUNDES2022110256} have given formulas for controllers for special systems with restrictions on the numbers and locations of right half plane poles and/or zeros. Some sophisticated numerical techniques have been brought to bear on some restricted problems in this area. Menini {\em et al.} \cite{Menini20231411} have assumed a fixed form for the controller along with bounds on coefficients therein, and examined whether a strongly stabilizing controller exists within those constraints. Niculescu and Michiels \cite{Niculescu2004802} have examined a particular form of the plant, namely one with poles at zero, and shown stabilization of the same with delayed feedback. These research papers show that a simple (as in accessible to a broad audience) yet general method for {\em computing} strongly stabilizing controllers is not yet available, five decades after Youla {\em et al.} \cite{youla1974single}.

This paper presents a new approach to the problem, under the restriction that the plant being stabilized should have relative degree less than three. In other words, our contribution is for plants wherein the degree of the denominator polynomial exceeds the degree of the numerator polynomial by 0, 1, or 2. We acknowledge that there are strong controllers given for plants of higher relative degree \cite{GUNDES2022110256}, but in those cases there are specific additional restrictions on the locations of poles and zeros of the open loop plant. Also, there is a sophisticated method based on Nevanlinna-Pick interpolation, about which we will write a little later. Our method only requires undergraduate level mathematics, and we place no further restrictions beyond requiring the relative degree to be less than three.

A key point in feedback stabilization, discussed in many papers including \cite{youla1974single}, is right half plane zeros of a controller are not allowed to cancel right half plane poles of the plant. Having said that, the parity interlacing property that is essential for strong stabilization to be possible is as follows.

\begin{dfn} \label{c1} {\em Parity interlacing property (PIP):
Between every pair of real non-negative zeros of the open loop plant, the number of real positive poles must be even.
Note that zeros at infinity count as real and positive; and absence of poles counts as an even number of poles (namely, 0).} \end{dfn}

 In theory, strong stabilizability does not depend on complex poles and zeros of the open loop plant. Nor does it depend on left half plane poles and zeros, whether real or complex.
In practice, complex conjugate pairs of right half plane zeros that are close to the real axis may lead to highly sensitive controllers of very high order, as noted in \cite{SMITH1986127, Hitayhinf}.

\begin{figure}[!htp]
    \centering
    \includegraphics[width = 0.48 \textwidth]{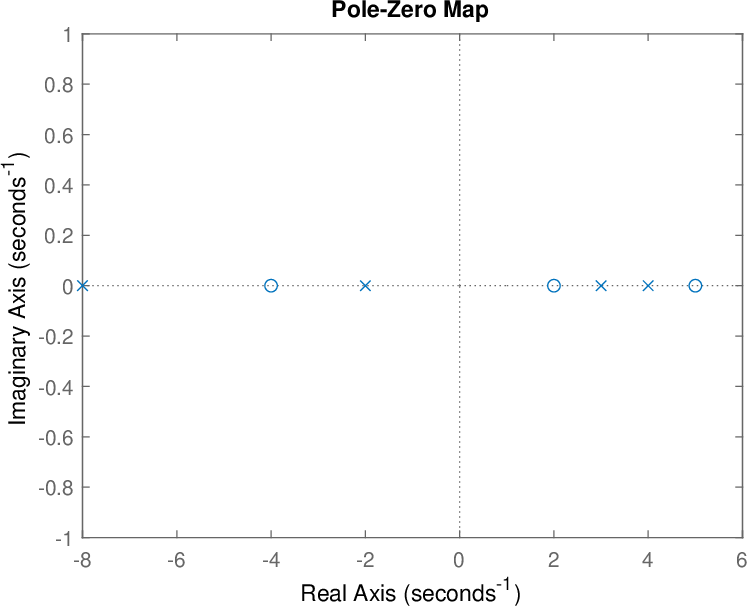}
    \caption{Pole-zero map of a hypothetical plant. Circles are zeros, and crosses are poles.}
    \label{fig1}
\end{figure}

To help fix ideas, consider the pole-zero map of a hypothetical plant given in Fig.\ \ref{fig1}. There are 4 poles and 3 zeros, so the plant is strictly proper (the relative degree is greater than 0, specifically 1). There are two finite positive zeros, and between them there is an even number of poles (namely two). There is also a zero at infinity, and between the rightmost positive finite zero and infinity there are no poles, which is even as well. Thus, this plant satisfies the PIP, and is strongly stabilizable. If the rightmost pole of this plant is relocated, e.g., to the left half plane or to the right of the rightmost zero, then the resulting new plant will not satisfy the PIP and will not be strongly stabilizable.
\bigskip

After checking that a plant is strongly stabilizable, it remains to compute a stabilizing stable controller. Here, things get specialized. Advanced textbooks on control theory do prescribe ways to find such controllers \cite{doyle2013feedback,vidyasagar2011control}, but the method has some arbitrariness because stabilizing controllers are not unique; and the formal steps involved are complicated as well (see, e.g., the 2018 paper \cite{Hitayhinf}).

In this paper we present a new numerical method for finding strong stabilizing controllers. The key idea is to use transfer functions with non-integer (real) powers in an intermediate step, and then adjust free parameters to turn those real powers into integer powers. For ease of reference, we refer to our method as {\em real to integer} (RTI). 

\section{Motivation for our approach}
Since our approach is unconventional, we now present its theoretical motivation.
We will use co-prime factorization and reduce our control design problem to considering four transfer functions that obey the relation
\begin{equation} \label{mg0} C(s) = \frac{U(s)-D(s)}{N(s)},
\end{equation}
where, if $N(s)$ has zeros in the right half plane, then $U(s)-D(s)$ must share those same zeros {\em exactly} so that a cancelation can occur and $C(s)$ has no right half plane poles. The reasoning for using the above formulation will be explained in Section \ref{bgt}. Here, we discuss how the above equation motivates our paper.

\begin{ex} \label{mg1} {\em As a simple example of Eq.\ (\ref{mg0}), we might have
\begin{equation} \label{zz1}
C(s) = \frac{ \displaystyle \frac{s+2}{s+5}  - \frac{s+5}{s+11} }  {\displaystyle \frac{s-1}{s+7}},
\end{equation}
where both numerator and denominator have a right half plane zero at $s=1$, and exact cancelation gives
\begin{equation} \label{zz2} C(s) = \frac{3(s+7)}{(s+5)(s+11)}. 
\end{equation}}
\end{ex}
Such simple and obvious cancelation, when it occurs in control design, is always welcome. However, as is well known within the topic of strong stabilization, sometimes such solutions are difficult to find. The starting point of our paper is the observation that cancelation of such zeros is in principle also possible if
$U(s)$ involves some non-integer, exponents or powers.

\begin{dfn} \label{fgs1} {\em Let $z = r e^{i \theta}$ be a complex number, with $r > 0$ and $-\pi < \theta \le \pi$. For $0 < \gamma < 1$, real powers of $z$ are defined to be
$$z^{\gamma} = r^{\gamma} e^{i \gamma \theta}.$$
The above definition implies a discontinuity on the negative real axis. The curve of discontinuity, in this case the negative real axis, is called a branch cut. Now let $N$ be any integer. We define
$$z^{N + \gamma} = z^N z^{\gamma},$$
where $z^N$ is routine and $z^{\gamma}$ is defined above.} \end{dfn}

\begin{dfn} \label{fgs2} {\em In complex analysis, the right half plane along with the imaginary axis is called the extended right half plane. However, for brevity, in this paper we will
say ``RHP'' to imply the extended right half plane.} \end{dfn}

Our approach is based on using a product of several terms of the form
\begin{equation} \label{FF} T = \left ( \frac{s+a}{s+b} \right )^m,
\end{equation}
where $a>0$, $b>0$, and $m$ is real. Since $m$ is allowed to be negative, there is no loss of generality in assuming $a<b$. In $T$ above, we can think of the complex number
$$z = \frac{s+a}{s+b},$$
where $s$ is any other complex number. Clearly, $z$ is real and negative only if $s$ is real and lies in $(-b,-a)$. That line segment is a branch cut for this function, and it lies strictly in the left half plane. Thus, $T$ is analytic for $s$ is in the RHP. $T$ also has no zeros in the RHP.

\begin{ex} \label{filp} {\em An example is now offered to demonstrate the meaningfulness of transfer functions involving fractions, as in Eq.\ (\ref{FF}).
Consider
\begin{equation} \label{gaw0} E(s) = \left ( \frac{s+3}{s+5} \right )^{2.7}.
\end{equation}
We begin by noting that the Dirac delta function $\delta(t)$ can appear in the inverse Laplace transforms of biproper plants. Consider
\begin{equation} \label{gaw1} {\cal L}^{-1} \left ( \frac{s+3}{s+5} \right ) = {\cal L}^{-1} \left (1 - \frac{2}{s+5} \right ) = \delta(t) - 2 e^{-5t}. \end{equation}
Now, we rewrite Eq.\ (\ref{gaw0}) as
$$ E(s) = \left ( \frac{s+3}{s+5} \right )^2 \cdot (s+3) \cdot \frac{1}{(s+3)^{0.3} } \cdot \frac{1}{(s+5)^{0.7}}.$$
The first factor is familiar, so we examine the reduced problem of
\begin{equation} \label{je1} H(s) = (s+3) \cdot \frac{1}{(s+3)^{0.3} } \cdot \frac{1}{(s+5)^{0.7}}\end{equation}
and its inverse Laplace transform.
Now the inverse Laplace transform of $s^{-0.7}$ is known:
$${\cal L}^{-1} \left ( \frac{1}{s^{0.7} } \right ) = \frac{1}{\Gamma (0.7) t^{0.3}},$$
where `$\Gamma$' refers to the gamma function. Thus,
$${\cal L}^{-1} \left ( \frac{1}{(s+5)^{0.7} } \right ) = \frac{e^{-5t}}{\Gamma (0.7) t^{0.3}}.$$
Similarly,
$${\cal L}^{-1} \left ( \frac{1}{(s+3)^{0.3} } \right ) = \frac{e^{-3t}}{\Gamma (0.3) t^{0.7}}.$$
The convolution of the above two functions gives the inverse Laplace transform of the fractional parts taken together, say
$$g(t) = {\cal L}^{-1} \left ( \frac{1}{(s+3)^{0.3} } \cdot \frac{1}{(s+5)^{0.7}} \right ) = {\cal L}^{-1} \left ( G(s) \right ) = \frac{1}{\Gamma (0.7)\Gamma (0.3)}
\int_0^t  \frac{e^{-3 \tau} e^{-5(t-\tau)}}{{\tau}^{0.7} (t-\tau)^{0.3}}  \, {\rm d} \tau.$$
Now $g(t)$ from such calculations generally cannot be expressed in terms of elementary functions (e.g., in this case
the symbolic algebra program Maple gives it in terms of the hypergeometric function). However, $g(t)$ is a well behaved, exponentially decaying function with $g(0^+) = 1$.
The Laplace transform of the derivative of $g(t)$ is
$${\cal L} \left ( \dot g(t) \right ) = sG(s) - g(0^+) = sG(s)-1.$$
Finally, we obtain
$$h(t) = {\cal L}^{-1} \left (H(s) \right ) =  \delta(t) + \dot g(t) + 3 g(t)$$
which, with the Dirac delta function present, is as acceptable as Eq.\ (\ref{gaw1}).
} \end{ex}

In spite of the foregoing example, real (non-integer) powers in transfer functions may still cause some readers to hesitate. To address their concerns, we offer the following comments. First, our real $m$'s appear only as intermediate steps in calculations; the final control design has integer powers only. In other words, the real $m$'s of our proposed solution exist in arithmetic and not physics.
 Also, note that even if we {\em did} have real $m$, there would exist a perfectly well defined inverse Laplace transform, as shown using the foregoing example.

We now demonstrate how real powers can be useful as intermediate steps in control design.

\begin{ex} \label{mg2} {\em Recall Eq.\ (\ref{zz1}). Now, for a simple example of a transfer function involving non-integer powers, consider
\begin{equation}
\label{ghm1} C(s) = \frac{ \displaystyle \left ( \frac{s+12}{s+19} \right )^{1.6090405507\cdots }  - \frac{s+5}{s+11} }  {\displaystyle \frac{s-1}{s+7}},
\end{equation}
which {\em also} has a canceling zero at $s=1$. The only difference is that now the cancelation cannot be carried out easily.} \end{ex}

\begin{ct} \label{mg3} {\em Using hardware
to realize separately the two parts,
namely
$$\left ( \frac{s+12}{s+19} \right )^{1.609\cdots} \mbox{ and } \frac{s+5}{s+11} ,$$
and then subtracting in hardware, will not yield an exact mathematical cancelation due to physical limitations and imperfections. Then the pole at $s=1$ will remain in $C(s)$, at least in principle.
In contrast, with integer exponents, {\em mathematical} cancelation can be carried out before physical realization, leading to a stable controller.} \end{ct}

\begin{ct} \label{c3} {\em The relevance of $T$ in Eq.\ (\ref{FF}) to Example \ref{mg2} is clear. Note, for later reference, that $T \rightarrow 1$ as $s \rightarrow \infty$.} \end{ct}

\begin{ex} \label{mg4} {\em Recalling Example \ref{mg2}, we now treat $b$ as a free parameter and start with the required cancelation at $s=1$ in
\begin{equation}
\label{ghm1i} C(s) = \frac{ \displaystyle \left ( \frac{s+12}{s+b} \right )^{1.60904055074\cdots }  - \frac{s+5}{s+11} }  {\displaystyle \frac{s-1}{s+7}}
\end{equation}
when $b=19$.
Now, if we continuously vary the parameter $b$, we can easily find that
\begin{equation}
\label{ghm2i} C(s) = \frac{ \displaystyle \left ( \frac{s+12}{s+17.38477631085 \cdots} \right )^{2 }  - \frac{s+5}{s+11} }  {\displaystyle \frac{s-1}{s+7}}
\end{equation}
{\em also} has a precisely canceling zero at $s=1$, where the real power has been {\em continuously} adjusted into an integer power.
Now we easily obtain (using Matlab's {\tt minreal} to numerically cancel the zero at $s-1$)
$$C(s) = - \frac{4.769552  s^2 +106.234631  s  + 509.934342}{s^3 + 45.769553 s^2 +   684.695526  s
+ 3324.534921 },$$
which is stable.
} \end{ex}

To summarize the motivation for our paper, in the strong stabilization problem we need to find a suitable transfer function
$U(s)$ that satisfies some properties. Allowing $U(s)$ to have real powers as above makes both satisfaction of those properties as well as computing the
$m$'s quite easy. Subsequently, adjusting free parameters to obtain integer $m$'s will also turn out to be easy. In contrast, doing the same thing in one shot, with predetermined integer $m$'s, wherein continuous variation of the powers is not possible, is a hard problem. It is in our adjustment process that we depart from existing methods.

\section{Background theory}
\label{bgt}
We now present a concise description of the background theory.  The material in this section is in principle well known. Nevertheless, the detailed explanation of why the strong stabilization problem is hard may be of interest to a broad audience.

A classical approach for strong stabilization was given by Youla {\em et al.} \cite{youla1974single}. More recently, the co-prime factorization approach \cite{vidyasagar2011control} has been adopted widely, and it is used here as well.

In co-prime factorization, the plant transfer function is expressed as
\begin{equation}
    P(s) = \frac{N(s)}{D(s)},
\end{equation}
where $N(s), \, D(s) \in \mathcal{S}$, by which we mean that $N(s)$ and $D(s)$ are themselves stable and proper transfer functions. Moreover, $N(s)$ and $D(s)$ are relatively co-prime, which means they have no common zeros ($s = \infty$ included). This implies $D(s)$ is {\em biproper}, which means its numerator and denominator polynomials are of the same degree. Finally, the zeros of $D(s)$ are restricted to RHP poles of $P(s)$. Note that co-prime factorization is not unique, but it is always possible.

\begin{ex} \label{exx1} Selection of $N(s)$ and $D(s)$. \end{ex}
Suppose the plant is
\begin{equation} \label{ex1}  P(s) = \frac{(s^2 - 3s + 7)(s+3)}{(s^2 + 4s + 8)(s-2)(s-3)}.
\end{equation}
This plant is strictly proper; it has relative degree one; and it has two zeros and two poles in the RHP. We can now use
\begin{equation} \label{ex2}
N(s) = \frac{s^2 - 3s + 7}{(s^2 + 4s + 8)(s+11)}, \quad D(s) = \frac{(s-2)(s-3)}{(s+3)(s+11)}.
\end{equation}

In the above, we note that the RHP zeros of $P(s)$ have remained as zeros of $N(s)$. The left half plane poles of $P(s)$ are now poles of $N(s)$. The left half plane zero of $P(s)$ is now a pole of $D(s)$. The RHP poles of $P(s)$ are zeros of $D(s)$. And the new arbitrary factor of $(s+11)$, introduced in the denominators of both $N(s)$ and $D(s)$, has a root in the left half plane and is used to make $D(s)$ biproper. Observing the steps in this example, it is clear that such a factorization is always possible for any plant whose transfer function is proper and a rational function of $s$. 

\begin{ct}
    \label{mcp}
    In the co-prime factorization of a plant $P(s)$ into $N(s)$ and $D(s)$, we could obviously multiply both
$N(s)$ and $D(s)$ by $-1$ to obtain another valid factorization. Except for one special situation described below, we will ensure that $\displaystyle \lim_{s \rightarrow \infty} D(s) = 1$. In the special case when $P(s)$ (i) is biproper,
        (ii) has at least one real RHP zero, and
        (iii) has an odd number of real poles to the right of its rightmost real zero,
we will multiply by $-1$ so that $\displaystyle \lim_{s \rightarrow \infty} D(s) = -1$.
\end{ct}

Let $C(s)$ be the stable controller transfer function. Define
\begin{equation}
\label{eq:CEm}
    R(s) = 1+P(s)\,C(s)  = 1 + \frac{N(S)}{D(s)}\,C(s).
\end{equation}
For closed-loop stability, $R(s)$ must have no RHP zeros. Note that $R(s)$ does have RHP poles. These poles are at the RHP zeros of $D(s)$, which are not allowed to cancel with zeros of either $N(s)$ or $C(s)$.

We now multiply by $D(s)$ and define
\begin{equation}
\label{bp1}
    U(s) = D(s) + N(s)\,C(s).
\end{equation}
Now $U(s)$ clearly does not have any RHP poles. This is because $N(s)$, $D(s)$ and $C(s)$ have no RHP poles. Since the closed-loop transfer function is
$$\frac{N(s)/D(s)}{1 + N(s)C(s)/D(s)} = \frac{N(s)}{D(s) + N(s)C(s)},$$
stability of that closed-loop system requires that $U(s)$ should not have any RHP zeros. A final, and crucial, condition remains from the control design viewpoint. If, instead of having a $C(s)$ and observing $U(s)$, we instead try to choose a $U(s)$ and back out the implied
$C(s)$, then we come to the previously introduced Eq. (\ref{mg0}),
\begin{equation}
\label{cd}
C(s) = \frac{U(s)-D(s)}{N(s)}.
\end{equation}

\begin{ct} \label{mgm0} {\em Stability of $C(s)$ above implies that the RHP zeros of $U(s)-D(s)$ include within them the RHP zeros of $N(s)$, counting zeros at infinity, and counting multiplicities.} \end{ct}

\begin{ct} \label{mgm1} {\em For the plant in Eq.\ (\ref{ex1}), both numerator and denominator polynomials are monic, i.e., their highest powers of $s$ have unit coefficients. There is no loss of generality because any non-unity constant can be absorbed into the controller, which is yet to be determined. In what follows we will assume that the numerator and denominator polynomials of the plant are, similarly, monic. Also, in Eq.\ (\ref{bp1}), since $D(s)$ is biproper, $U(s)$ is assumed to be biproper as well.} \end{ct} 

Given $N(s)$ and $D(s)$ as above, the design of a strongly stabilizing controller for $P(s) = N(s)/D(s)$ thus reduces to the choice of a biproper $U(s)$ which has three properties:
\begin{enumerate}
\item $U(s)$ is stable; 
\item $1/U(s)$ is stable also; and
\item $U(s)-D(s)$ has the RHP zeros of $N(s)$, counting both infinity and multiplicities.
\end{enumerate}

For a general reader, it may be useful to briefly consider some naive approaches that may come to mind. Let us (naively)
assume that $U(s)$ is a rational function of $s$, of the form
\begin{equation}
\label{uform}
U(s) = \frac{\alpha_0 + \alpha_1 s + \alpha_2 s^2 + \cdots + \alpha_n s^n}{1 + \beta_1 s + \beta_2 s^2 + \cdots + \beta_n s^n},
\end{equation}
where the numerator and denominator degrees are assumed to be the same because $U(s)$ has been taken to be biproper, where the 1 in the denominator ensures that there is no pole at 0 (which is in the RHP), and where the
$\alpha$'s and $\beta$'s are parameters to be chosen.
Generally, we do not know what $n$ needs to be. For some systems, it is known that $n$ must be quite large. If we assume an $n$ that is not large enough, there is no solution.

\begin{ct} \label{c8} {\em Zeros at infinity need care. If $N(s)$ has relative degree $k$, then $U(s)-D(s)$ must have a zero of degree at least
$k$ at infinity. With increasing $k$, this places increasingly complicated constraints on the $\alpha$'s and $\beta$'s of Eq.\ (\ref{uform}).
If $D(s)$ is chosen to be monic, then choosing $\alpha_n=\beta_n$ in Eq.\ (\ref{uform}), or otherwise ensuring that $U(s) \rightarrow 1$ as
$s \rightarrow \infty$, takes care of $k=0$ and $k=1$. In this connection recall the large-$s$ behavior of $T$ noted in Remark \ref{c3}.} \end{ct}

Having chosen $n$ and matched zeros at infinity, we still have an interpolation problem. $U(s)$ must match $D(s)$, i.e., $U(s)-D(s)$ must have zeros, at all of the finite RHP zeros of $N(s)$, counting multiplicities.

In the special case where $N(s)$ has no finite RHP zeros, this interpolation problem is avoided. Some examples of this important special case are considered
in \cite{GUNDES2022110256}.

We now come to the main interpolation problem. For simplicity, let us consider the case where the RHP zeros of $N(s)$ all have multiplicity one. We thus need
\begin{equation}
\label{Uzi}
U(z_i) = D(z_i),
\end{equation}
where the $z$'s are the RHP zeros of $N(s)$. Subject to the constraints of Remark \ref{c8}, we must now assign values to the $\alpha$'s and $\beta$'s such that the Eqs.\ (\ref{Uzi}) are satisfied. The challenge lies in the further constraint that all the roots of the numerator {\em and} of the denominator in Eq.\ (\ref{uform}) must lie in the left half plane.

 Unfortunately, for many problems, simple attempts at interpolation based on directly using Eqs.\ (\ref{Uzi}) to fit the coefficients in Eq.\ (\ref{uform}) give RHP zeros either for the numerator or denominator or both. Such solutions are not acceptable.

Given the difficulties outlined above, an alternative (also naive) approach might be to assume the form
\begin{equation}
\label{uform1}
U(s) = \frac{(s+\alpha_1)^{r_1}(s+\alpha_2)^{r_2} \cdots (s+\alpha_n)^{r_n}}{(s+\beta_1)^{r_{n+1}}(s+\beta_2)^{r_{n+2}} \cdots (s+\beta_n)^{r_{2n}}},
\end{equation}
where the $r$'s are integers and the $\alpha$'s and $\beta$'s are positive. Unfortunately, for given choices of the $r$'s, there may either be no strictly positive solutions for the $\alpha$'s and $\beta$'s, or they may be difficult to determine due to the high-dimensional and nonlinear equations arising from the interpolation problem of Eq.\ (\ref{Uzi}).

It is now clear that the difficulty in choosing $U(s)$ lies in that it must interpolate between the RHP zeros of $N(s)$, {\em and that} it must have neither poles nor zeros in the RHP. That  problem can be addressed by an advanced technique called Nevanlinna-Pick interpolation, which could be called the state of the art, and which we avoid in this paper.  The relevant point here is that the Nevanlinna-Pick interpolant is bounded and has no poles in some region of interest, but it could in principle have zeros there; and so additional tricks are needed, as described next.

We close this section with a brief sketch of one modern existing approach that uses Nevanlinna-Pick interpolation for the case where the RHP zeros of $N(s)$ all have multiplicity one \cite{Hitayhinf}. The approach is based on assuming
$$U(s) = F(s)^\ell, \quad \| F \|_{\infty} < 1, \quad \ell \ge 1 \mbox{ is an integer},$$
satisfying
Eq.\ (\ref{Uzi}),
where
$$F(s) = \frac{\displaystyle G(s)+1}{\displaystyle G(s)+\rho},  \quad \| G \|_{\infty} < 1, \quad \rho > 1 \, \mbox{ to be chosen,}$$
and where, finally, $G(s)$ is constructed using Nevanlinna-Pick interpolation \cite{HitayNP}. The integer $\ell$, if chosen large enough, guarantees a solution \cite{Hitayhinf}.

\begin{ct} \label{c13} {\em In the above approach, an additional search is performed in the $\rho$-$\ell$ space to minimize $\ell$, in order to reduce the degree of the corresponding controller. We will do a similar search to reduce the degrees of our controllers in our own method below.} \end{ct}

A more direct approach than Nevanlinna-Pick but also a more limited one, involving an iterative construction that is not easy to generalize to situations with many RHP zeros, is given in textbooks \cite{doyle2013feedback,vidyasagar2011control}. This direct approach is an adaptation of the method originally proposed by Youla {\em et al.} \cite{youla1974single}. We do not address that approach here because of its known limitations.

The foregoing discussion clarifies why strong stabilization is indeed a hard problem. In what follows, we bring
a fresh perspective to this problem and present a general solution approach for systems whose relative degree is less than three. The restriction on relative degree is because of zeros at infinity (recall Remark \ref{c8}).

\section{Our new theory}
\label{secAFP}
Our approach has two distinct parts. In the first part, we assign a form to $U(s)$ which includes free parameters and guarantees the essential conditions of stability and no RHP zeros. The price paid is that $U(s)$ includes a product of terms of the form of Eq.\ (\ref{FF}), wherein the $m$'s are typically real. Details of the plant determine the number of factors in $U(s)$ in a straightforward way, and the different possibilities will be discussed one by one below. In the second part of the approach, we first adjust the free parameters to make the $m$'s small overall, and then adjust them again to turn them into integers. Here we use a simple trigonometric trick: $m$ is an integer if and only if $\sin m \pi = 0$. With this trick, we can vary the free parameters and use continuous optimization and root finding methods, as opposed to discrete search.

\subsection{The form of $U(s)$}
\label{someL}
Recall Eq.\  (\ref{cd}). Note that the numerator $U(s)-D(s)$ is allowed to have additional zeros in the RHP, beyond canceling the zeros of $N(s)$.

\begin{ct} \label{kn1} {\em In Eq.\  (\ref{cd}), if $N(s)$ is biproper, i.e., its relative degree is zero, then it does not have a zero at infinity. If $N(s)$ has relative degree one, then it has a single zero at infinity. Both cases are handled if we ensure that
$U(s)-D(s)$ has a zero of multiplicity one at infinity.} \end{ct}

To this end, note that
$$ \lim_{s \rightarrow \infty} D(s) = 1$$
because $D(s)$ is monic. So we require
$$ \lim_{s \rightarrow \infty} U(s) = 1.$$
Recalling Eq.\ (\ref{FF}) and also Remark \ref{c8}, the above requirement is obviously met if
$U(s)$ is taken to be a product of a finite number of terms of the form
$$ \left ( \frac{s+a}{s+b} \right )^m.$$

Relative degree two
will be addressed later, and it will be seen that higher degrees can in principle also be accommodated, but analytical complexities will increase. For this reason, in this paper, we restrict ourselves to plants with relative degree less than three.

\begin{ct} \label{kn2} {\em By Eq.\ (\ref{cd}), our control design task ends with our choice of $U(s)$. For the numerical realization of $C(s)$, the cancellation between the zeros of $U(s)-D(s)$ and those of $N(s)$ must be actively carried out. If implemented in Matlab, the built-in command \texttt{minreal} may be used to compute this minimal realization of the controller (recall Example \ref{mg4}).} \end{ct}

Our $U(s)$ will always have $2r$ strictly positive free parameters, for which we will begin with
$$0 < a_1 < a_2 < \cdots < a_{2r}.$$

Note that, if the implied ordering in the $a$'s above is changed after adjustment of parameters, it will have no consequence.

Given the $a$'s, we define
\begin{equation}
\label{eq:fk}
f_k = \frac{s+a_{2k-1}}{s+a_{2k}}.
\end{equation}
Then we choose the form
\begin{equation}
\label{formu}
U(s) = \prod_{k=1}^r f_k^{m_k},
\end{equation}
where the indices $m_k$ are to be determined.

\begin{ct} \label{kn3} {\em Finding $U(s)$ contains within itself an interpolation problem, as discussed above. Interpolation through
distinct points to match given values is simplest. Matching derivatives at some points adds equations including derivatives of the interpolant.
But it need not change the mathematical form of the interpolant. In particular, the form of our interpolant guarantees that it has neither RHP zeros nor RHP poles. This eliminates the need for Nevanlinna-Pick interpolation.} \end{ct}

Further details of $U(s)$ depend on details of the plant. The different possibilities are now discussed one by one.

\subsection{Relative degree zero, with all RHP zeros simple}
\label{zzx}
In this case, note that the order of the plant is not being restricted. Since there is no zero at infinity, we do not require $U(s)-D(s)$ to be zero at infinity. We must check if Remark \ref{mcp} is applicable.

We begin by counting the number of {\em
finite} RHP zeros of $N(s)$. Let this number be $q$.\\
If $q=0$, there is nothing left to do. Choosing
\begin{equation}
\label{zzx1}
U(s) = 1
\end{equation}
will yield a stable stabilizing controller. No non-integer powers will arise.

For biproper $P(s)$ with $q>0$, let the number of real poles of $P(s)$ the right of its rightmost real positive zero be $q_p$. If $q_p$ is odd, we use the $-1$ multiplier mentioned in Remark \ref{mcp}. Then $\displaystyle \lim_{s\rightarrow \infty}D=-1$; and 
$D(s)$ changes signs an odd number of times on the real line; so that at the rightmost zero of $N(s)$, we have $D(s) > 0$; and subsequently, because of the PIP, at every real positive zero of $N(s)$, we have $D(s) > 0$. The positive sign of $D(s)$ at the real RHP zeros of $N(s)$ allows us to work with logarithms, which is a key part of our method. If $q_p$ is even, we avoid the $-1$ multiplier for the same reasons.

Beyond this small but important step of possibly multiplying by $-1$, plants with relative degree zero are handled the same as plants with relative degree one, which we discuss next.

\subsection{Relative degree one, with all RHP zeros simple}

The zero at infinity has already been taken care of (recall Remark \ref{kn1}). For $q=0$, we can choose $U(s)=1$. Non-integer powers do not arise.

We now proceed with $q > 0$. We use $r=q$ in Eq.\ (\ref{formu}) and choose
$$0 < a_1 < a_2 < \cdots < a_{2q}.$$
These choices are arbitrary at this stage, and will be numerically adjusted using procedures described shortly.

Let the $q$ finite RHP zeros of $N(s)$ be $z_1, z_2, \cdots, z_q.$
As explained above, we require
$$U \left (z_n \right ) = D \left (z_n \right ), \quad 1 \le n \le q.$$
Taking logarithms, we have
\begin{equation}
\label{cd1}
\sum_{k=1}^q m_k \ln \left [ f_k \left (z_n \right ) \right ] = \ln \left [ D \left (z_n \right ) \right ], \quad 1 \le n \le q.
\end{equation}
Equations (\ref{cd1}) are linear in the indices $m_k$, and can be solved directly. It is important to check if solutions will be real.

\begin{pp} \label{pp1} {\em Solutions for the $m_k$ will be real if strong stabilization is possible.} \end{pp}

\noindent {\bf Proof:} For complex zeros of $N(s)$, the equations will appear in complex conjugate pairs, allowing real solutions. For real RHP zeros of $N(s)$, our selective use of the $-1$ multiplier (recall Remark \ref{mcp}) ensures that $D(s)$ is positive at all such zeros if the plant satisfies the PIP.
Thus, logarithms cause no difficulties and real solutions occur. $\blacksquare$

\begin{ct} \label{kn5} {\em Having obtained real $m$'s, the corresponding $U(s)$ can technically give us a stable stabilizing controller (recall
Examples \ref{filp} and \ref{mg2}). However, real $m$'s in $U(s)$ cause difficulty in the subtraction and cancellation required in Eq.\ (\ref{cd}). This is why we seek integer solutions for the $m$'s, so that cancellations become routine, as noted in
Remark \ref{kn2}.} \end{ct} 

We now establish that integer solutions are possible for the $m$'s determined using Eq.\ (\ref{cd1}).

\subsection{Existence of integer solutions}
\label{proof}
We will adjust the $a$'s to obtain integer $m$'s. To show that integer $m$'s are possible, we will
first use an asymptotic calculation to establish the intuitive idea, and then formalize it using the implicit function theorem. The implied
integer solutions are large, but they demonstrate existence.

Recalling
\begin{equation} \label{pf}
f_k^{m_k} = \left ( \frac{s+a_{2k-1}}{s+a_{2k}} \right )^{m_k},
\end{equation}
we assign
\begin{equation}
\label{smallsep} a_{2k-1} = b_k-\epsilon_k, \mbox{ and } a_{2k} = b_k+ \epsilon_k, \quad 0 < |\epsilon_k| \ll 1.
\end{equation}
Taking logarithms of both sides of Eq.\ (\ref{pf}) for use in Eq.\ (\ref{cd1}), we have
\begin{equation}
\label{eq:asy}
 m_k \ln f_k = m_k \ln \left ( 1 - \frac{2\,\epsilon_k}{s+b_k+\epsilon_k} \right ) = \left ( -   \frac{2\,\epsilon_k}{s+b_k} -
\frac{2\,\epsilon_k^3}{3(s+b_k)^3} + \cdots \right ) \cdot m_k.
\end{equation}
Now we write
\begin{equation}
\label{limsol}
x_k = 2 \epsilon_k m_k,
\end{equation}
and obtain
\begin{equation}
\label{IFT1}
m_k \ln f_k = m_k \ln \left ( 1 - \frac{2\,\epsilon_k}{s+b_k+\epsilon_k} \right ) = \left ( -   \frac{1}{s+b_k} -
\frac{\epsilon_k^2}{3(s+b_k)^3} + \cdots \right ) \cdot x_k.
\end{equation}
For intuition, we can drop the ${\cal O} \left (\epsilon_k^2 \right )$ terms above and plug the approximation into Eq.\ (\ref{cd1}), to obtain a linear system of the form
\begin{equation}
\label{asy}
\mathbf{Ax} = \mathbf{b},
\end{equation}
where
$\mathbf{A}$, $\mathbf{x}$, and $\mathbf{b}$ are independent of the $\epsilon$'s. Upon obtaining $x_k$ in this way, we can choose a sufficiently large integer value $N_k$, and obtain an asymptotic estimate of the corresponding design parameter,
\begin{equation}
\label{asy1}\epsilon_k = \frac{x_k}{2 N_k}.
\end{equation}
Very small adjustments of the $\epsilon$'s in Eq.\ (\ref{smallsep}), inserted in Eq.\ (\ref{cd1}), then give solutions
$m_k$ that are equal to the arbitrarily pre-specified large integers $N_k$. Since large integers are also integers, this suggests strongly that integer solutions exist. 
A numerical example is  given in Appendix \ref{app:proof}. 
We will present a formal proof below.

For ease of presentation below, we recall the implicit function theorem (see e.g., \cite{rudin1953principles}).
Consider an $n_1$-dimensional vector valued function $\mathbf{f}(\boldsymbol{\eta}, \boldsymbol{\xi})$, of vectors $\boldsymbol{\eta}, \boldsymbol{\xi}$ which are $n_1$-dimensional and $n_2$-dimensional respectively. 
If $\mathbf{f(\boldsymbol{\eta^*}, \boldsymbol{\xi^*})}=\mathbf{0}$, and the $n_1 \times n_1$ Jacobian matrix  of $\mathbf{f}$ with respect to $\boldsymbol{\eta}$ at $(\boldsymbol{\eta^*}, \boldsymbol{\xi^*})$ is invertible, then
there exists an implicit functional form
$
\boldsymbol{\eta} = \mathbf{g(\boldsymbol{\xi})}
$
which satisfies
$\mathbf{f({g\boldsymbol(\boldsymbol{\xi})}, \boldsymbol{\xi})}=\mathbf{0},$
whereby $\boldsymbol{\eta}$ is uniquely determined for every $\boldsymbol{\xi}$ in some neighbourhood of $\boldsymbol{\xi^*}$.

\begin{dfn} \label{df1} {\em Let $\mathbf{D_m}$ denote the diagonal matrix with diagonal elements $m_1, m_2, \cdots, m_q$. Also, let
$\mathbf{D}_{\mathbf{m}}^{-1}$ denote the diagonal matrix with diagonal elements $1/m_1, 1/m_2, \cdots, 1/m_q$.} \end{dfn}

\begin{dfn} \label{df2} {\em Let $\boldsymbol{\epsilon}$ denote the column matrix whose elements are $\{ \epsilon_1, \epsilon_2,
\cdots, \epsilon_q \}$.} \end{dfn}

\begin{pp} \label{tm2} {\em If there are choices of positive $a$'s for which the matrix {\bf A} in Eq.\ (\ref{asy}) is invertible, then there are
integer solutions for the $m$'s in Eq.\ (\ref{cd1}).} \end{pp}

\noindent {\bf Proof:} 
Recalling Eqs.\ (\ref{cd1}), (\ref{IFT1}) and (\ref{asy}), we write
$$2 (\mathbf{A} + \mathbf{B}) \mathbf{D_m} \boldsymbol{\epsilon} = \mathbf{b},$$
where $\mathbf{A}$ is assumed invertible as before; and where $\mathbf{B}$ is
${\cal O} \left ( \| \boldsymbol{\epsilon} \|^2 \right )$.

We thus obtain
$$ \boldsymbol{\epsilon} - \frac{1}{2} \mathbf{D}_{\mathbf{m}}^{-1} \left ( \mathbf{I} + \mathbf{A}^{-1}\mathbf{B} \right )^{-1} \mathbf{A}^{-1} \mathbf{b} = \mathbf{0}.$$
For sufficiently small $\boldsymbol{\epsilon}$, $\mathbf{A}^{-1}\mathbf{B}$ is small as well (e.g., in the induced 2-norm), so we have the convergent series
$$ \left ( \mathbf{I} + \mathbf{A}^{-1}\mathbf{B} \right )^{-1} = \left ( \mathbf{I} - \mathbf{A}^{-1}\mathbf{B} + \left ( \mathbf{A}^{-1}\mathbf{B} \right )^2 \cdots \right ),$$
which is only ${\cal O} \left ( \| \boldsymbol{\epsilon} \|^2 \right )$ different from the identity matrix.
Thus,
\begin{equation} \label{ift} \boldsymbol{\epsilon} - \frac{1}{2} \mathbf{D}_{\mathbf{m}}^{-1} \left ( \mathbf{I} + {\cal O} \left ( \| \boldsymbol{\epsilon} \|^2 \right ) \right ) \mathbf{A}^{-1} \mathbf{b} = 0.
\end{equation}
In Eq.\ (\ref{ift}), our variables are in vector $\boldsymbol{\epsilon}$, and our parameters are the reciprocals of the indices, i.e., $1/m_k$,
in $\mathbf{D}_{\mathbf{m}}^{-1}$. The Jacobian with respect to $\boldsymbol{\epsilon}$ is only ${\cal O} \left ( \| \boldsymbol{\epsilon} \| \right )$ different from the identity matrix, and is exactly equal to the identity matrix when $\boldsymbol{\epsilon}$ is zero. Clearly, Eq.\ (\ref{ift}) is satisfied if the $1/m$'s are all zero, and $\boldsymbol{\epsilon}$ is zero as well. Thus, by the implicit function theorem, for arbitrary but sufficiently small values of the $1/m$'s, there exist unique solutions for the $\epsilon$'s. In particular, the small values of the $1/m$'s may obviously be chosen as the reciprocals of large enough, {\em but otherwise arbitrary}, integers. This proves that integer solutions exist. $\blacksquare$

Although we have formally proved the existence of integer solutions that might be very large, numerical examples
below will demonstrate that integer solutions of moderate size typically exist.

We now turn to computing integer solutions of moderate size using a trigonometric trick.

\subsection{Seeking integer solutions of smaller size}
Having established that integer solutions exist, we 
wish to adjust our parameters to obtain integer solutions of small to moderate size.
We do this in three steps.

First, to achieve smallness, we run an optimization step where we adjust the
$a$'s to obtain small $m$'s, possibly non-integer. This is a reasonable step: recall Remark \ref{c13}.

Second, to nudge the $m$'s towards integer values, we run a second optimization step wherein we use a trigonometric trick.

\begin{ct} \label{q2} {\em Our trigonometric trick is that $m_k$ is an integer if and only if $\sin m_k \pi = 0.$} \end{ct}

Accordingly, in the second step, we minimize
$$J=\sum_{k=1}^q \sin^2 \left ( m_k  \pi \right ),$$
where the objective function $J$ achieves zero whenever every $m$ is an integer. If we obtain very small values of $J$, we have essentially obtained
integer $m$'s.

\begin{ct} \label{q3} {\em In the above optimization, we have incorporated an additional constraint, requiring each of the $a$'s to be greater than unity. Note that strictly positive $a$'s are all that we require. The constraint is an added arbitrary step to yield better numerical conditioning. For example, $(s+1)^{10} = 0$ has 10 roots exactly equal to $-1$. However, $(s+1)^{10} = 10^{-10}$ has 10 nonzero roots that differ from $-1$ by 0.1 in magnitude. Similar perturbations of $(s+0.05)^{10}=0$ would give roots in the RHP.} \end{ct}

Third, to refine the result of our second optimization to high accuracy (e.g., several decimal points), we
run a modified Newton-Raphson iteration to take the vector valued quantity
$$\sin \left ({\bf m}  \pi \right )$$
to zero.

These three steps were implemented in Matlab. Details are given in Appendix \ref{app:AFP}.

So far we have considered $N(s)$ whose RHP zeros have multiplicity one. We now turn to finite RHP zeros of higher multiplicity. It will be seen that only small changes are needed in the approach already described.

\subsection{Repeated finite RHP zeros}
We now consider $N(s)$ that has RHP zeros of multiplicity greater than one, except that the zero at infinity (if any) still has multiplicity one. In other words, the relative degree of the plant is still less than two.

To present the idea in a simple way, we first consider the case where {\em one} zero has multiplicity two. Now, one of the equations in Eq.\ (\ref{cd1}) is redundant. Suppose there are $q$ finite RHP zeros, and 
$z_{q-1} = z_q$. Then we only have $q-1$ independent equations so far, i.e., 
\begin{equation}
\label{cd1m}
\sum_{k=1}^q m_k \ln \left [ f_k \left (z_n \right ) \right ] = \ln \left [ D \left (z_n \right ) \right ], \quad 1 \le n \le q-1,
\end{equation}
wherein we have $q-1$ distinct $z$'s and $q$ distinct $f$'s.

In Eq.\ (\ref{cd1m}), $z_q$ yields the same equation as $z_{q-1}$, and so that equation adds nothing.
The $q^{\rm th}$ equation is obtained by using derivatives of both sides for the $(q-1)^{\rm th}$ equation. In general, multiplicity $j+1$ requires $j$ successive derivatives. Accordingly, for the case of multiplicity 2, we have the final equation, also linear:
\begin{equation}
\label{eder}
\sum_{k=1}^q m_k  \frac{f_k' \left (z_{q-1} \right )}{f_k \left (z_{q-1} \right )}  = \frac{ D' \left (z_{q-1} \right ) }{D \left (z_{q-1} \right ) }.
\end{equation}
Solution for the $m$'s is straightforward, and the same adjustment procedures for the parameters can be used to obtain integer $m$'s.

Clearly, if there is a complex conjugate pair of roots with multiplicity two, then there will be $q-2$ distinct $z$'s, $q$ distinct $f$'s,
$q-2$ independent equations in a further-reduced version of Eq.\ (\ref{cd1m}), and two equations involving first order derivatives in place of the single Eq.\ (\ref{eder}). The extension of the theory to such cases is obvious and therefore not discussed. Examples will follow.

We finally turn to plants with relative degree higher than 1.

\subsection{$P(s)$ with higher relative degree}
\label{hrd}

Suppose the relative degree of $P(s)$ is $k>0$. This means $N(s)$ has a zero of multiplicity $k$ at infinity. This in turn means that
$U(s)-D(s)$ must have a zero of multiplicity at least $k$ at infinity. To that end, we present the following almost obvious theorem.

\begin{pp} \label{t3} {\em $U(s)-D(s)$ has a zero of multiplicity $k$ at infinity if the large-$s$ expansion of $U(s)-D(s)$, in inverse powers of $s$, has
$s^{-k}$ in its leading term.} \end{pp}

\noindent {\bf Proof:} $D(s)$ is monic, so it has some {\em finite} poles but is analytic for large enough $s$. Dividing its numerator and denominator by the highest power of $s$ present, we obtain an equivalent expression involving inverse powers of $s$. We then write
$$\sigma = \frac{1}{s},$$
and it is clear that $D=1$ when $\sigma=0$; and that $D$ has a Taylor series expansion about $\sigma=0$. Next,
$U(s)$ is composed of a number of multiplying
factors of the form
$$\left ( \frac{s+a_{2k-1}}{s+a_{2k}} \right )^{m_k}$$
with $m_k$ real. The above factor can be rewritten as
$$\displaystyle \left ( \frac{1+\displaystyle \frac{a_{2k-1}}{s}}{1+\displaystyle \frac{a_{2k}}{s}} \right )^{m_k} = \left ( \frac{1+{a_{2k-1}}{\sigma}}{1+{a_{2k}}{\sigma}} \right )^{m_k}.$$
The right hand side above shows that each such factor is also equal to unity when $\sigma=0$, and also that each such factor has a straightforward Taylor series expansion about $\sigma=0$ even if $m_k$ is real. Writing the quantity $U-D$ in terms of $\sigma$ in this way, and expanding the result
in a Taylor series for small $\sigma$, we obtain something of the form
$$ \alpha_0 + \alpha_1 \sigma + \alpha_2 \sigma^2 + \cdots.$$
If $\alpha_0 \neq 0$, there cannot be a zero at infinity. However, since both $U(s)$ and $D(s)$ approach unity as $s \rightarrow \infty$, we
do have a zero, and so we know that $\alpha_0=0$. Similarly, if $\alpha_1 \neq 0$, we have a zero of multiplicity one at $s = \infty$. And so on for higher multiplicities. Thus, if $U(s)-D(s)$ has a zero of multiplicity $k$ at infinity, then 
$$\alpha_k \sigma^k = \frac{\alpha_k}{s^k}$$
must be the first nonzero term in the expansion above.
$\blacksquare$

\begin{ct} \label{q6} {\em In particular, if $U(s)-D(s)$ has a zero of multiplicity 2 at infinity, then the $s^{-1}$ term in its large-$s$ expansion must have zero coefficient.} \end{ct}

We now consider the case where the relative degree of $P(s)$ is 2.
First, we expand $D(s)$ in a series using inverse powers of $s$. We need only the $s^{-1}$ term in that series. To ensure a zero of degree 2 at infinity for $U(s)-D(s)$, the $s^{-1}$ term in a similar series for $U(s)$ must be equal to the one arising from $D(s)$.

As before, let $q$ be the number of finite RHP zeros of $N(s)$. If $q=0$, then choosing $U(s)$ is simplest.
Let
\begin{equation}
\label{ro2a}
D(s) = \frac{s^p + b_1 s^{p-1} + \cdots}{s^p + c_1 s^{p-1} + \cdots},
\end{equation}
then for large $s$,
\begin{equation}
\label{ro2b}
D(s) = 1 + \frac{b_1-c_1}{s} + \cdots.
\end{equation}

Therefore, for plants with relative degree 2 and $q=0$, we pick any number $M >0$ such that $b_1 - c_1 + M > 0$. Then we use
\begin{equation}
\label{cd1b}
U(s) = \frac{s + b_1 - c_1 + M}{s+M},
\end{equation}
and we have a stable stabilizing $C(s)$. Non-integer powers do not arise.

We now consider $q>0$. Let the finite RHP zeros of $N(s)$ be 
$z_1, z_2, \cdots, z_q$ as before.

We first consider the case where the RHP zeros of $N(s)$ all have multiplicity one.
In this case we obtain a useful $U(s)$ by choosing $2q+2$ positive parameters
\begin{equation} \label{upars} 0 < a_1 < a_2 < \cdots < a_{2q+2},
\end{equation}
and use
\begin{equation}
\label{cd1a} U(s) = \frac{s + b_1 - c_1 + M}{s+M} \, \prod_{k=1}^{q+1} f_k^{m_k},
\end{equation}
where
$f_k$, $b_1$, $c_1$ and $M$ are as defined already.
Now there are $q+1$ indices $m_k$ to be determined. We have $q$ equations from taking logarithms, as follows:
\begin{equation}
\label{cd2}
\sum_{k=1}^{q+1} m_k \ln f_k \left (z_n \right ) = \ln D \left (z_n \right ) - \ln \left ( \frac{z_n + b_1 - c_1 + M}{z_n+M} \right ), \quad 1 \le n \le q.
\end{equation}

\begin{ct} \label{q9} {\em The correspondence between Eqs.\ (\ref{cd2}) and (\ref{cd1}) is clear. In particular, only the $m$'s are considered unknowns because the other parameters including $M$ are chosen by us. We  are still one equation short because we have
$q$ equations and there are $q+1$ $m$'s.} \end{ct}

\begin{pp} \label{t4} {\em Equation (\ref{cd2}) must be complemented with
the linear equation
\begin{equation}
\label{cd3}
\sum_{k=1}^{q+1} m_k \left ( a_{2k-1}-a_{2k} \right ) = 0.
\end{equation}
The resulting $m$'s will typically be real but will give, using Eq. \eqref{cd}, a stable stabilizing $C(s)$.} \end{pp}

\noindent {\bf Proof:} By Theorem \ref{t3} and Remark \ref{q6}, the $s^{-1}$ term in a large-$s$ expansion of $U(s)-D(s)$ needs to be zero.  Equations (\ref{ro2b}) and (\ref{cd1b}) already present a design wherein the $s^{-1}$ terms cancel. Equations (\ref{cd1b}) and
(\ref{cd1a}) are identical except for the multiplier
$$\prod_{k=1}^{q+1} f_k^{m_k},$$
whose $s^{-1}$ term in a large-$s$ expansion is zero if and only if Eq.\ (\ref{cd3}) is satisfied. $\blacksquare$

The extension to RHP zeros of multiplicity two is straightforward. As before, we lose one equation from Eqs.\ (\ref{cd2}), and add one equation involving derivatives, like Eq.\ (\ref{eder}). Higher multiplicities require higher derivatives, in a straightforward way. Note that Eq.\ (\ref{cd3}) is not affected.

\begin{ct} \label{q11} {\em Finally, the real (non-integer) powers are to be adjusted to integer values by varying the free parameters in $U(s)$
(recall Eq.\ (\ref{upars}). The preceding theory for such adjustment is directly applicable.} \end{ct}

This concludes our theory. Theorem \ref{t3} indicates the additional conditions to be met if plants of relative degree greater than 2 are to be included in this formulation. We leave that to future work because the constraints on the parameters become nonlinear, simple explicit linear equations for determining the $m$'s are not available, and so it complicates the presentation.

\section{Examples}
We now present several examples, of different types of systems, to demonstrate the performance of our proposed RTI approach. For completeness, some graphical representations of the controlled systems' step input responses are given at the end of this section.
Below, for brevity, we will refer to the parameter vector containing $a$'s as $\mathbf{a}$.

\begin{enumerate}
    \item \textbf{$\mathbf{P(s)}$ has relative degree 0}

\begin{ex}  \label{exx3} Relative degree zero, and $q=1$. \end{ex}   
        Consider
    \begin{equation}
        P(s) = \frac{(s-3)\,(s+2)}{(s-4)\,(s-5)}.
    \end{equation}
    We choose
    \begin{align*}
        N(s) &= \frac{(s-3)}{(s+3)},\\
        D(s) &= \frac{(s-4)\,(s-5)}{(s+2)\,(s+3)}.
    \end{align*}
Choosing $\mathbf{a} = [1, 17]^T$ as an arbitrary parameter vector, we obtain $m = 1.68261$. As described in Appendix \ref{app:AFP}, we adjust the parameters to obtain integer $m$'s. The results are given in Table \ref{tab:exx3}.
\begin{table}[h]
    \centering
    \begin{tabular}{||r|r|r|r||}
    \hline
    Initial parameter & Initial indices &  Adjusted parameters & $m$'s obtained \\ 
    choice ($a$'s) & ($m$'s) & ($a$'s) & using adjusted $a$'s\\
    \hline
    1  & 1.68261 & 1.000000000 & 1\\
    17 &         & 57.00000000 & \\
    \hline
    \end{tabular}
    \caption{Parameters ($a$'s) and exponential indices ($m$'s) for Example \ref{exx3}.}
    \label{tab:exx3}
\end{table}

    That the parameter $a_1$ is exactly unity is a consequence of an {\em ad hoc} decision explained in Appendix \ref{appb1}. This pattern will repeat itself
in examples below. With the adjusted parameters as given above, we obtain
    \begin{equation}
        U(s) = \frac{s+1}{s+57}.
    \end{equation}
    The resulting controller strongly stabilizes the system. $\blacksquare$

\item $\mathbf{P(s)}$ \textbf{has relative degree 1}

\begin{ex} \label{exx2} Relative degree one, and $q=0$. \end{ex}
Consider
$$P(s) = \frac{s+1}{s^2 - s + 5}.$$
We choose
$$N(s) = \frac{s+1}{s^2 + s + 5} \quad \mbox{ and } \quad D(s) = \frac{s^2 - s +5}{s^2 + s +5}.$$
In the above, it is clear that the denominator polynomial in $D(s)$ just needs to be of degree 2 and stable.
As stated earlier (recall Eq.\ (\ref{zzx1}), Section \ref{zzx}), we can choose $U(s)=1$.
Equation (\ref{cd}) gives
$$C(s) = \frac{2s}{s+1}.$$
The closed-loop transfer function,
$$\frac{N(s)}{D(s)+N(s)C(s)},$$
can be easily computed and is indeed stable. $\blacksquare$

\begin{ex}  \label{exx4} Earlier example, now completed. \end{ex}
    Consider again the plant in example \ref{exx1}, with the same $N(s)$ and $D(s)$, i.e.
    \begin{equation} \label{ex4}  
        P(s) = \frac{(s^2 - 3s + 7)(s+3)}{(s^2 + 4s + 8)(s-2)(s-3)},
    \end{equation}
    and
    \begin{align*}
    N(s) &= \frac{s^2 - 3s + 7}{(s^2 + 4s + 8)(s+11)}, \\
    D(s) &= \frac{(s-2)(s-3)}{(s+3)(s+11)}.
    \end{align*}
Results obtained are given in Table \ref{tab:exx4}.

\begin{table}[h]
    \centering
    \begin{tabular}{||r|r|r|r||}
    \hline
    Initial parameter & Initial indices &  Adjusted parameters & $m$'s obtained \\ 
    choice ($a$'s) & ($m$'s) & ($a$'s) & using adjusted $a$'s\\
    \hline
    10  &  $-27.9055$ & 1.000000000  & $-7$\\
    37  &             & 8.565360692  & \\
    82  &    63.4279  & 12.05378853  & 4\\
    145 &             & 178.9280213  & \\
    \hline
    \end{tabular}
    \caption{Parameters ($a$'s) and exponential indices ($m$'s) for Example \ref{exx4}.}
    \label{tab:exx4}
\end{table}

    Thus, we obtain
    \begin{equation}
        U(s) = \left(\frac{s+8.565360692}{s+1}\right)^7\left(\frac{s+12.05378853}{s+178.92802131}\right)^4.
    \end{equation}
    The corresponding controller is strongly stabilizing.
$\blacksquare$

\begin{ex} \label{exx5} Relative degree one, $q=2$. \end{ex}
    Consider 
    \begin{equation}
        P(s)=\frac{s^2 - 2s + 5}{(s-2.5)\, (s^2 + 2s + 5)}.
    \end{equation}
    We choose
    \begin{align*}
        N(s) &= \frac{s^2 - 2s + 5}{(s+2.5)\, (s^2 + 2s + 5)},\\
        D(s) &= \frac{(s-2.5)}{(s+2.5)}.
    \end{align*}

\begin{table}[h]
    \centering
    \begin{tabular}{||r|r|r|r||}
    \hline
    Initial parameter & Initial indices &  Adjusted parameters & $m$'s obtained \\ 
    choice ($a$'s) & ($m$'s) & ($a$'s) & using adjusted $a$'s\\
    \hline
    5  &  $5.8606$      & 1.000000000 & $3$\\
    101 &               & 12.65454035 & \\
    226 & $-11.4661$    & 14.62249082 & $-2$\\
    901 &               & 132.6597271 & \\
    \hline
    \end{tabular}
    \caption{Parameters ($a$'s) and exponential indices ($m$'s) for Example \ref{exx5}.}
    \label{tab:exx5}
\end{table}

    Results are shown in Table \ref{tab:exx5}. We thus obtain
    \begin{equation}
       U(s) = \left(\frac{s+1.000000000}{s+12.65454035}\right)^3\left(\frac{s+132.6597271}{s+14.62249082}\right)^2.
    \end{equation}
    The controller thus obtained strongly stabilizes the system. $\blacksquare$

\item $\mathbf{P(s)}$ \textbf{has repeated RHP zeros}
\begin{ex} \label{exx6} Real zero with multiplicity $2$, $q=2$ (counting multiplicities).\end{ex}
 Consider 
    \begin{equation}
        P(s)=\frac{(s-2)^2}{(s+6)\,(s-3) \,(s-4)}.
    \end{equation}
    We choose
    \begin{align*}
        N(s) &= \frac{(s-2)^2}{(s+6)\,(s+3) \,(s+4)},\\
        D(s) &= \frac{(s-3)\,(s-4)}{(s+3) \,(s+4)}.
    \end{align*}

Results are shown in Table \ref{tab:exx6}.

\begin{table}[h]
    \centering
    \begin{tabular}{||r|r|r|r||}
    \hline
    Initial parameter & Initial indices &  Adjusted parameters & $m$'s obtained \\ 
    choice ($a$'s) & ($m$'s) & ($a$'s) & using adjusted $a$'s\\
    \hline
    5   &  $-14.7788$   & 1.000000000 & $-9$\\
    101 &               & 9.207908073 & \\
    226 & $30.8386$     & 12.31517239 & $5$\\
    901 &               & 261.8400886 & \\
    \hline
    \end{tabular}
    \caption{Parameters ($a$'s) and exponential indices ($m$'s) for Example \ref{exx6}.}
    \label{tab:exx6}
\end{table}

    Thus,
    \begin{equation}
       U(s) = \left(\frac{s+9.207908073}{s+1.000000000}\right)^9\left(\frac{s+12.31517239}{s+261.8400886}\right)^5.
    \end{equation}
    The resulting controller strongly stabilizes the system. $\blacksquare$

\begin{ex} \label{exx7} Complex conjugate zeros with multiplicities $2$, $q=4$.\end{ex}

 Consider 
    \begin{equation}
        P(s)=\frac{ (s^2 - 4s + 40)^2}{(s+2)\,(s+6)\,(s+8)\,(s+10)\,(s-4)}.
    \end{equation}
    We choose
    \begin{align*}
        N(s) &= \frac{ (s^2 - 4s + 40)^2}{(s+2)\,(s+6)\,(s+8)\,(s+10)\,(s+4)},\\
        D(s) &= \frac{(s-4)}{(s+4)}.
    \end{align*}
\begin{table}[h]
    \centering
    \begin{tabular}{||r|r|r|r||}
    \hline
    Initial parameter & Initial indices &  Adjusted parameters & $m$'s obtained \\ 
    choice ($a$'s) & ($m$'s) & ($a$'s) & using adjusted $a$'s\\
    \hline
    1.01  & $186.9702$    & 1.000000000 & $12$\\
    1.09  &               & 3.125685736 & \\
    1.81  & $-2.7053$     & 3.020123314 & $-7$\\
    8.29  &               & 11.00083916 & \\
    66.61 & $5.4911$      & 13.14342623 & $5$\\
    577   &               & 67.80945410 & \\
    5185  & $-5.0108$     & 383.9773935 & $3$\\
    46657 &               & 77.84899459 & \\
    \hline
    \end{tabular}
    \caption{Parameters ($a$'s) and exponential indices ($m$'s) for Example \ref{exx7}.}
    \label{tab:exx7}
\end{table}    
    
    Results obtained are given in Table \ref{tab:exx7}. Note that, after adjustment, the parameters $a_k$ are not monotonically increasing. That does not matter, because we have stabilized the
system. The monotonically increasing nature of $a$'s was useful to establish the existence of integer solutions. Thus, we obtain
    \begin{equation}
       U(s) = \left(\frac{s+1.000000000}{s+3.125685736}\right)^{12}\left(\frac{s+11.00083916}{s+3.020123314}\right)^7\left(\frac{s+13.14342623}{s+67.80945410}\right)^5\left(\frac{s+383.9773935}{s+77.84899459}\right)^3.
    \end{equation}
    The controller thus obtained strongly stabilizes the system. $\blacksquare$

\item $\mathbf{P(s)}$ \textbf{has relative degree 2}
\begin{ex} \label{exx8a} Relative degree $2$, $q=0$. \end{ex}
Consider
$$P(s) = \frac{s+1}{(s^2-s+4)(s+3)}.$$
We let
\begin{align*}
N(s) &= \frac{s+1}{(s^2 + s + 7)(s+3)}, \\
D(s) &= \frac{s^2 - s +4}{s^2 + s + 7}.
\end{align*}
In terms of Eqs.\ (\ref{ro2a}) and (\ref{ro2b}),
$$b_1 - c_1 = -2.$$
We choose $M=3$. That yields
$$U(s) = \frac{s+1}{s+3},$$
which in turn yields the stabilizing controller
$$C(s) = \frac{7s-5}{s+1},$$
which is stable. $\blacksquare$

\begin{ex} \label{exx8} Relative degree two, $q=2$. \end{ex}
    Consider
    \begin{equation}
        P(s) = \frac{(s-5) (s-2)}{(s-3) (s-4) (s+2.5) (s+1.5)}.
    \end{equation}
    We choose
    \begin{align*}
        N(s) &= \frac{(s-5) (s-2)}{(s+3) (s+4) (s+2.5) (s+1.5)},\\
        D(s) &= \frac{(s-3) (s-4)}{(s+3) (s+4)} = \frac{ s^2 - 7 s + 12}{ s^2 + 7 s + 12}.
    \end{align*}
    Hence, $b_1= -7, c_1 = 7$, and $b_1-c_1 = -14$; we choose $M = 15$. Thus, the pre-multiplier term becomes
    $$
    U_p(s) = \frac{s+1}{s+15}.
    $$

\begin{table}[h]
    \centering
    \begin{tabular}{||r|r|r|r||}
    \hline
    Initial parameter & Initial indices &  Adjusted parameters & $m$'s obtained \\ 
    choice ($a$'s) & ($m$'s) & ($a$'s) & using adjusted $a$'s\\
    \hline
    2     & $-4.4306$     & 1.000000000 & $-5$\\
    10    &               & 8.488509423 & \\
    82    & $2.7321$      & 9.252626592 & $4$\\
    730   &               & 94.36909940 & \\
    6562  & $-0.0340$     & 405.8562852 & $1$\\
    57601 &               & 102.8329410 & \\
    \hline
    \end{tabular}
    \caption{Parameters ($a$'s) and exponential indices ($m$'s) for Example \ref{exx8}.}
    \label{tab:exx8}
\end{table} 

    Results obtained are shown in Table \ref{tab:exx8}. Thus,
    \begin{equation}
    U(s) = \left(\frac{s+1}{s+15}\right)\left(\frac{s+8.488509423}{s+1}\right)^5\left(\frac{s+9.252626592}{s+94.36909940}\right)^4\left(\frac{s+405.8562852}{s+102.8329410}\right).
    \end{equation}
    The resulting controller strongly stabilizes the system. $\blacksquare$

\item $\mathbf{P(s)}$ \textbf{has relative degree 2 with repeated zeros}
\begin{ex} \label{exx9} Relative degree two with repeated complex conjugate zeros, $q=4$. \end{ex}
    Consider
    \begin{equation}
        P(s) = \frac{(s^2 - 4s + 40)^2}{(s-4)\,(s+2)\,(s+6)\,(s+8)\,(s+10)\,(s+12)}.
    \end{equation}
    We choose
    \begin{align*}
        N(s) &= \frac{(s^2 - 4s + 40)^2}{(s-4)\,(s+2)\,(s+6)\,(s+8)\,(s+10)\,(s+12)},\\
        D(s) &= \frac{s-4}{s+4}.
    \end{align*}
    Upon choice of $M=9$ using calculations as shown above, the pre-multiplier is
    $$
    U_p(s) = \frac{s+1}{s+9}.
    $$

\begin{table}[h]
    \centering
    \begin{tabular}{||r|r|r|r||}
    \hline
    Initial parameter & Initial indices &  Adjusted parameters & $m$'s obtained \\ 
    choice ($a$'s) & ($m$'s) & ($a$'s) & using adjusted $a$'s\\
    \hline
    1   & $3.6973$      & 1.000006671 & $12$\\
    5   &               & 2.936514430 & \\
    17  & $-38.9268$    & 2.664991202 & $-7$\\
    37  &               & 241.2744419 & \\
    65  & $319.3825$    & 12.86646544 & $13$\\
    101 &               & 78.89989125 & \\
    145 & $-601.2791$   & 64.17384002 & $-1$\\
    197 &               & 210.3103283 & \\
    257 & $301.9484$    & 221.8268170 & $2$\\
    325 &               & 689.1918246 & \\
    \hline
    \end{tabular}
    \caption{Parameters ($a$'s) and exponential indices ($m$'s) for Example \ref{exx9}.}
    \label{tab:exx9}
\end{table}    

    We observe again, from Table \ref{tab:exx9}, that the $a$'s are not increasing monotonically as originally assumed, and that it does not matter. Thus,
    \begin{align}
    \nonumber
    U(s) &= \left(\frac{s+1}{s+9}\right)\left(\frac{s+1.000006671}{s+2.936514430}\right)^{12}\left(\frac{s+241.2744419}{s+2.664991202}\right)^7\left(\frac{s+12.86646544}{s+78.89989125}\right)^{13}\\&\left(\frac{s+210.3103283}{s+64.17384002}\right)\left(\frac{s+221.8268170}{s+689.1918246}\right)^{2}.
    \end{align}
    The controller corresponding to this $U(s)$ strongly stabilizes the system. $\blacksquare$
\end{enumerate}

\begin{figure}[!htp]
    \centering
    \includegraphics[width = 14 cm]{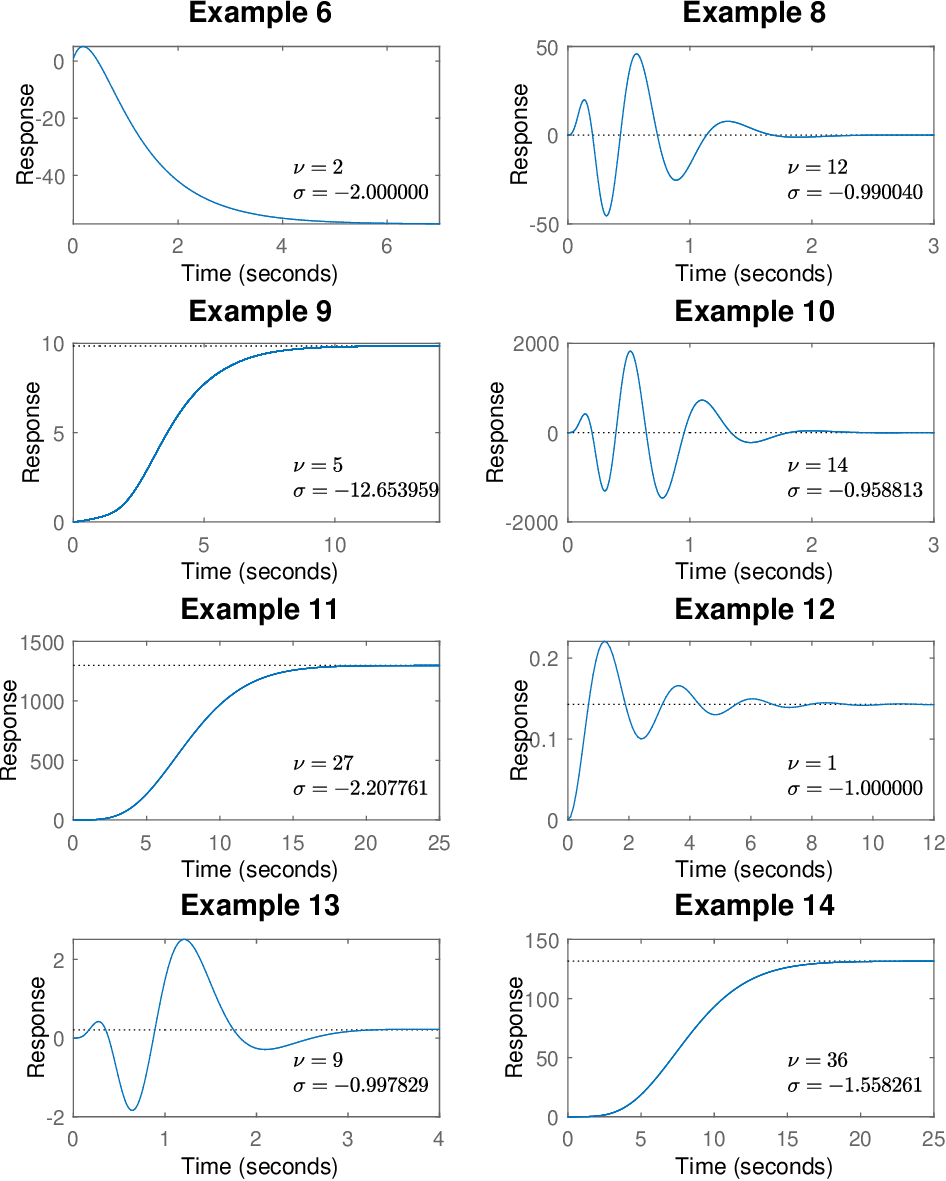}
    \caption{Closed-loop step responses for examples \ref{exx3} through \ref{exx9} (example \ref{exx2} is trivial and excluded). In each subplot, $\nu$ is the degree of the denominator polynomial of the controller, and $\sigma$ is the largest of the real parts of the controller poles. In order to achieve the cancellations implied by Eq.\ (\ref{cd}) using Matlab's {\tt minreal}, error tolerances had to be adjusted in some cases, for reasons explained in Remark \ref{q3}.}
    \label{fig:steps}
\end{figure}

In the interest of providing further useful detail, the step responses of the controlled systems from eight of the foregoing examples are shown in Fig.\ \ref{fig:steps}.
In each subplot, $\nu$ denotes the order of the denominator polynomial of the controller $C(s)$. Also given is $\sigma$, the largest of the real parts of the controller poles: in each case, $\sigma<0$ shows that the controller is stable. Finally, the fact that the step response settles down at a steady finite value shows that the system has indeed been stabilized. Thus, strong stabilization was achieved in every case.

\section{Conclusions}
In classical linear time-invariant single-input single-output control with a single feedback loop, the problem of strong stabilization has been of sustained interest. While modern control theory, which uses state space methods, is important in computer-based systems, there are still applications where a simple
analog card used in the feedback loop can be used fruitfully. In such cases, there are operational advantages if the controller itself is stable.

Widely available methods in textbooks, beyond the co-prime factorization, present controller design using manual search. Such methods are difficult to apply to somewhat complicated systems. A more modern approach uses Nevanlinna-Pick interpolation, which is an advanced mathematical technique. That design approach works solely with controllers of integer order.

In this work, we have presented a significant departure from the usual approaches, in that we have started with a transfer function that is guaranteed to be stable, with a stable inverse, with an adequate number of free parameters, and which allows real powers in intermediate calculations. The advantage of allowing real powers is that the control design is reduced to numerical solution of some linear simultaneous equations which determine those real powers. Subsequently, any of many possible {\em ad hoc} parameter adjustment methods can be used so that the final powers used in the controller are integer powers after all. We have presented an informal asymptotic argument explaining why such integer powers can always be found, formally proved the same using the implicit function theorem, and finally demonstrated the same with a numerical example.
Further, we have used robust though {\em ad hoc} ways in which the free parameters can be adjusted to obtain such integer solutions. Finally, we have presented 10 numerical examples worked out in detail, with the plant's relative degree being 0, 1 or 2; both with and without RHP zeros of the plant; and with such zeros possibly having multiplicity greater than one. Our algorithm has worked successfully in all cases.

We believe that, in terms of both higher simplicity in the algorithm and lower mathematical sophistication needed for its implementation, our proposed method is new. There is no comparable method in the literature. Extension to more difficult cases not presently included, namely relative degrees of 3 or higher, are left to future work. 

\section*{Acknowledgements}
We thank Hitay \"Ozbay for sharing and discussing his papers, lecture notes, as well as Matlab code for the Nevanlinna-Pick interpolation based controller design approach. We thank C.P. Vyasarayani and Andy Ruina for editorial comments, and Dennis Bernstein for technical inputs on some related problems. 

In an earlier version of this article, we had referred to our method as {\em adjustment of fractional powers} (AFP), but have now decided to use the term RTI to avoid confusion with treatments that actually use fractional order calculus.

\bibliography{citations}

\section*{Appendices}
\appendix
\section{Existence of integer-power solutions}
\label{app:proof}
We now provide a numerical example to accompany our discussion (Section \ref{proof}) of the existence of possibly large integer solutions for the $m$'s.

\begin{ex} Solutions that are large pre-specified integers. \end{ex}
Consider 
\begin{equation}
    P(s) = \frac{ (s^2 - 4.1s + 5.9) (s^2 - 2.6s + 5.3)}{(s-5.8) (s^2 + 4.1s + 5.9) (s^2 + 2.6s + 5.3)},
\end{equation}
where we have chosen the numerical parameters arbitrarily. We choose\footnote{%
	Note that $\pi = 3.1\underline{{\bf 4159265358}}97 \cdots$. The underlined digits were used in this example.}
\begin{align*}
    N(s) &= \frac{ (s^2 - 4.1s + 5.9) (s^2 - 2.6s + 5.3)}{(s+5.8)(s^2 + 4.1s + 5.9) (s^2 + 2.6s + 5.3)},\\
    D(s) &= \frac{s-5.8}{s+5.8}.
\end{align*}
We require 8 parameters. Following Eq.\ (\ref{smallsep}), let these be of the form
\begin{equation}
\label{pv}
{\bf a} = 
    \begin{bmatrix}
    a_1\\
    a_2\\
    a_3\\
    a_4\\
    a_5\\
    a_6\\
    a_7\\
    a_8
    \end{bmatrix} = 
    \begin{bmatrix}
    b_1-\epsilon_1\\
    b_1+\epsilon_1\\
    b_2-\epsilon_2\\
    b_2+\epsilon_2\\
    b_3-\epsilon_3\\
    b_3+\epsilon_3\\
    b_4-\epsilon_4\\
    b_4+\epsilon_4
    \end{bmatrix}.
\end{equation}
We choose the $b$'s arbitrarily to be $b_1 =2, b_2 =4, b_3 = 8, b_4 = 16$.
The asymptotic Eq.\ (\ref{asy}) yields
\begin{equation}
\label{asy2}
    \mathbf{x} = \begin{bmatrix}
        598.6609674630\\
        -3220.323300969\\
        7469.405659544\\
        -6450.310943517
    \end{bmatrix}.
\end{equation}
Now, we arbitrarily decide that the $\epsilon$'s should be about 0.05 in magnitude (a small number), and assign $N_1 = 6000, N_2 = 32000, N_3 = 75000, N_4 = 65000$,
which yield, from Eqs.\ (\ref{asy2}) and (\ref{asy1}),
\begin{align*}
    \epsilon_1&= 0.049888413955255\\
    \epsilon_2&= -0.050317551577652 \quad .\\
    \epsilon_3&= 0.049796037730294\\
    \epsilon_4&= -0.049617776488599
\end{align*}
Upon substituting the above $\epsilon$'s in Eq.\ (\ref{pv}), and solving the exact Eq.\ (\ref{cd1}),
we obtain
$$m_1 = 5995.358742, m_2 = 31978.54572, m_3 = 74958.13240, m_4 = 64968.66745.$$
These are close to, though not equal to, the specified integer values. 
We can now adjust the $\epsilon$'s very slightly to obtain the desired integer indices. Results are given in Table \ref{tab:comp}. $\blacksquare$
\begin{table}[h]
\vspace{0.1in}
    \centering
    \begin{tabular}{||r|r|r|r||}
    \hline
    Pre-specified & Predicted $\epsilon$'s by asymptotic& Final $\epsilon$'s obtained & $m$'s corresponding\\
    integers ($N$) &  approximation  &  by adjustment &  to the adjusted $\epsilon$'s\\ \hline
        6000  &  0.049888413955255 &  0.049849885843271 & 6000.0000000\\
        32000 & $-0.050317551577652$ & $-0.050283869393989$ & 32000.0000000\\
        75000 &  0.049796037730294 &  0.049768282530780 & 75000.0000000\\
        65000 & $-0.049617776488599$ & $-0.049593895055261$ & 65000.0000000\\
        \hline
    \end{tabular}
    \caption{The $\epsilon$'s obtained from the asymptotic approximation can be adjusted slightly to yield the integer
powers specified. The numerical match is good to 7 decimal places, as shown.}
    \label{tab:comp}
\vspace{0.1in}
\end{table}

Thus, sufficiently small $\epsilon_k$ can always be chosen to ensure that each $m_k$ is an integer: in fact, infinitely many such large integer solutions can be {\em specified}. In practice, as shown by earlier examples, integer solutions of smaller size can usually be found by numerical search.

\section{Implementation of the RTI method}
\label{app:AFP}
\subsection{Search for smaller integer solutions using two-stage optimization}
\label{appb1}
For any parameter vector $\mathbf{a}$, we can compute a set of indices $m_k$. Now we can define the preconditioning
objective function
\begin{equation}
\label{F0}
F_0(\mathbf{a})=\sum_{k=1}^q |m_k|,
\end{equation}
and minimize it with respect to ${\bf a}$ to get the indices down to smaller values. At this stage the indices remain real (possibly non-integer).

Note that, instead of the sum of absolute values, we could have used the sum of squares as well, or any of many other possible objective functions. The choice is 
{\em ad hoc} and the controller is non-unique.

As a practical matter, we have taken three steps to make the preconditioning optimization robust. First, for unconstrained input $\tilde a_k$,
we specified
\begin{equation}
\label{nlm}
a_k = 1 + \tilde a_k^2,
\end{equation}
so that optimization searches do not wander into regions with negative $a$'s. Second, we penalized extremely large values of the $a$'s, to make sure that optimization searches do not wander off to infinity. Finally, we penalized any $m_k$ approaching too-small values, in case the zero solution (i.e., $m_k=0$) is not possible for the problem at hand. Thus, mapping $\tilde a$'s to $a$'s as in Eq.\ (\ref{nlm}), we adjusted Eq.\ (\ref{F0}) and actually minimized
$$F_1(\mathbf{a})=\sum_{k=1}^q |m_k|+Q,$$
where 
$$Q = 10  \left ( 1- \min_k |m_k| \right ) \cdot H \left ( 1- \min_k |m_k| \right ) + 0.01 \sum_n a_n^2,$$
where in turn $H$ denotes the Heaviside function ($H(x) = 1$ if $x>0$, and $H(x)=0$ otherwise) and where the coefficients 10 and 0.01 (large and small, respectively) are arbitrarily chosen.

After lowering the magnitudes of the indices $m_k$, we nudge them towards integer values by a second optimization calculation using the
objective function
$$F_2(\mathbf{a}) = \sum_{k=1}^q \sin^2 \left ( m_k  \pi \right ),$$
where again we use $a$'s from Eq.\ (\ref{nlm}). In the above, it is clear that $F_2 \ge 0$, with $F_2 = 0$ only when every $m_k$ is an integer.

The two optimization calculations above are unconstrained and do not require sophisticated algorithms. We have used Matlab's built-in {\tt fminsearch}. At the end of these optimization calculations, we have obtained $m$'s that are quite close to integers (typically up to 4 or more decimal places). We now carry out a further refinement using modified Newton-Raphson iterations.

\subsection{Modified Newton-Raphson}
\label{appb2}
We now define the $q$-dimensional vector function
$$\mathbf{g}(\mathbf{\tilde a}) = \sin \left ( \mathbf{m} \pi \right ),$$
where the tilde is used to denote the unconstrained variables introduced using Eq.\ (\ref{nlm}).
We will iteratively adjust the $\tilde a$'s until $\mathbf{g}$ is acceptably close to zero. We note that, due to the prior optimization, we are starting with $\mathbf{g}$ quite close to zero, and so convergence is practically assured. Now, the input vector has $2q$ elements while $\mathbf{g}$ has only $q$ elements. The Newton-Raphson iteration requires a Jacobian or matrix of first partial derivatives. We numerically estimate the Jacobian using finite differences. The Jacobian is not a square matrix: it is $2q \times q$.

The Newton-Raphson iteration seeks $\Delta \mathbf{\tilde a}$ such that
\begin{equation}\label{NR} \mathbf{g}(\mathbf{\tilde a}) + \mathbf{J} \cdot \Delta \mathbf{\tilde a} = \mathbf{0},
\end{equation}
where $\mathbf{J}$ is the Jacobian evaluated at the current $\mathbf{\tilde a}$.

Here, $\mathbf{J}$ is not square, and Eq.\ (\ref{NR}) is under-determined. So we introduce a $q$-dimensional vector of Lagrange multipliers, and minimize $(\Delta \mathbf{\tilde a})^T (\Delta \mathbf{\tilde a})$ subject to the constraint given by Eq.\ (\ref{NR}). This is a routine problem in matrix algebra, outlined here for completeness.

In order to minimize
$$\frac{\mathbf{y}^T\mathbf{y}}{2} \mbox{ subject to } \mathbf{By} = \mathbf{c},$$
where matrix $\mathbf{B}$ has more columns than rows,
we solve
\begin{equation}
\label{minN} \left [ \begin{array}{cc} \mathbf{I} & \mathbf{B}^T \\ \mathbf{B} & \mathbf{0} \end{array} \right ] \left \{ \begin{array}{c} \mathbf{y} \\ \boldsymbol{\lambda} \end{array} \right \}
= \left \{ \begin{array}{c} \mathbf{0} \\ \mathbf{c} \end{array} \right \},
\end{equation}
where $\mathbf{I}$ and $\mathbf{0}$ denote identity and zero matrices of appropriate sizes, and where the Lagrange multiplier vector $\boldsymbol{\lambda}$ can be discarded after the calculation.

With the above modified Newton-Raphson iteration, we have obtained excellent results, including every example given in the previous section. A minor technical point here is that if one column of the rectangular matrix $\mathbf{B}$ happens to be zero but $\mathbf{B}$ is of full rank, then the solution of Eq.\ (\ref{minN}) remains valid.
This is relevant in our case because one of the $\tilde a$'s becomes zero for many examples (the corresponding $a$ taking the value 1).

\end{document}